\documentclass[11pt,amstex]{article}
\usepackage{amsmath,amssymb,amsmath,amsfonts,amsthm}
\usepackage{fancyhdr}
\begin{document}
\title{Calculation of generalized Hubbell rectangular source integral}

\author{Jonathan Murley and Nasser Saad\footnote{Corresponding author} \\ \\
Department of Mathematics and Statistics, University of Prince Edward Island\\
Charlottetown, Prince Edward Island  C1A 4P3, Canada\\ \\
{\bf E-Mail: nsaad@upei.ca}\\
}
\date{\today}

\maketitle

\begin{abstract}
\noindent A simple formula for computing
 the generalized Hubbell radiation rectangular source integral
$$H\left[
{a,b,p,\lambda \atop
\alpha,\beta,\gamma }  
          \right]=\frac{\sigma a}{4\pi}\int_0^b x^\lambda(x^2+p)^{-\alpha}{}_2F_1(\alpha,\beta;\gamma;-{a^2\over x^2+p})dx,
$$
is introduced. Tables are given to compare the numerical values derived from our approximation formula  with those given earlier in the literature. 
\end{abstract}

   \vskip0.1true in                        
\noindent{\bf keywords:} {~Appell hypergeometric functions $F_2$, Hubbell radiation rectangular source integral, Classical hypergeometric functions, Reduction formulas, Radiation field integrals.}

\section{Introduction}

In their pioneering work, Hubbell et al. (1960)  obtained a series expansion for the calculation of radiation field generated by a plane isotropic rectangular source (plaque), in which the leading term is the integral
\begin{align}\label{sec1eq1}
I(a,b)&={\sigma\over 4\pi} \int_0^b \arctan\left({a\over \sqrt{x^2+1}}\right){dx\over \sqrt{x^2+1}}, 
\end{align}
here $\sigma$ is the uniform surface source strength per unit source area. In equation (\ref{sec1eq1}) the quantities $a=w/h$ and $b=l/h$ are defined in the range $0<a\leq b\leq\infty$, where $h$ is the height over the a corner of a plaque of length $l$ and width $w$.  For the important applications of this integral in many problems in radiation field, different  methods were introduced to obtain numerical values of detector response to plaque source
\begin{equation}\label{sec1eq2}
h(a,b)={I(a,b)\over ({\sigma/4\pi})}= \int_0^b \arctan\left({a\over \sqrt{x^2+1}}\right){dx\over \sqrt{x^2+1}}
\end{equation}
For some of these methods we refer to the work of Glasser, 1984; Kalla et al., 1987; Ghose et al., 1988; Galu\'e et al., 1988; G\"otze, 1995; Kalla, 1993; Timus, 1993; Michieli and Maximovic, 1996; Kalla and Khajah (1997, 2000); Prabha, 2001; Stalker, 2001;  Guseinov et al., 2004; Ezure, 2005; Guseinov et al., 2005; Prabha, 2007. 
\vskip0.1true in
Although $h(a,b)$ is not expressible in simple closed form, Glasser (1984) has evaluated it in terms of Appell's hypergeometric function $F_2$ (see for example Slater (1966), Ch. 8, for a study of Appell functions $F_q,~q=1,2,3,4$). Indeed, elementary differentiation of (\ref{sec1eq2}), with respect to $a$, we have
\begin{equation}\label{sec1eq3}
h(a,b)=\int_0^a dx \int_0^b {dy\over {1+x^2+y^2}}
\end{equation}
and straightforward substitutions $x=a\sqrt{u}$ and $y=b\sqrt{v}$ allow as to write (\ref{sec1eq3}) as the double-integral representation of Appell's hypergeometric function $F_2$ (Slater (1966), Ch. 8, formula 8.2.3), consequently
 \begin{equation}\label{sec1eq4}
h(a,b)=ab F_2(1;{1\over 2},{1\over 2};{3\over 2},{3\over 2};-a^2,-b^2).
\end{equation}
Various generalizations of equation (\ref{sec1eq1}) have been given in the literature  (see for example, Kalla et al., 1987;  Galu\'e et al., 1988; Saigo and Srivastava, 1990; Galu\'e, 1991; Kalla, 1993; Galu\'e et al., 1994, Kalla et al. 2002, Oner 2007).  More specifically, Kalla et al. (1987) introduced a generalization defined by the integral
\begin{equation}\label{sec1eq5}
H\left[\begin{array}{l}
a,b,p,\lambda    \\
\alpha,\beta,\gamma   
          \end{array}\right]=\frac{\sigma a}{4\pi}\int_0^b x^\lambda(x^2+p)^{-\alpha}{}_2F_1(\alpha,\beta;\gamma;-{a^2\over x^2+p})dx
\end{equation}
where $\gamma>\beta>0;a,b,p>0;-1<\lambda<2a-1;$ and ${}_2F_1(\alpha,\beta;\gamma;x)$ is Gauss hypergeometric function (Slater, 1966, Ch. 1). We notice that
\begin{equation}\label{sec1eq6}
H\left[\begin{array}{l}
a,b,1,0    \\
1,\frac{1}{2},\frac{3}{2}   
          \end{array}\right]=I(a,b)
\end{equation}
by virtue of the identity (Slater, 1966; formula 1.5.11)
\begin{equation}\label{sec1eq7}
x{}_2F_1(1,{1\over 2};{3\over 2};-x^2)=\arctan(x).
\end{equation}
By selecting suitable values for the parameters $\alpha,\beta$ and $\gamma$, equation (\ref{sec1eq5}) can be reduced to different integrals with potential applications in radiation-field problems of specific configurations of source, barrier and detector (Kalla, 1993). Such results are also useful in illumination and heat-exchange engineering Boast, 1942; Fano et al., 1959; Hubbell, 1960. Using a simple transformation,  $x=b\sqrt{u}$, equation (\ref{sec1eq5}) can be written as
\begin{equation}\label{sec1eq8}
H\left[\begin{array}{l}
a,b,p,\lambda    \\
\alpha,\beta,\gamma   
          \end{array}\right]=\frac{\sigma a}{4\pi}{b^{\lambda+1}\over 2p^\alpha} \int_0^1 u^{{\lambda+1\over 2}-1} (1-u)^{{\lambda+3\over 2} -{\lambda+1\over 2}-1} (1-{-b^2u\over p})^{-\alpha}{}_2F_1(\alpha,\beta;\gamma;{-{a^2\over p}\over 1-(-{b^2\over p})u})du
\end{equation}
which is easily compared with the single-integral representation of the Appell hypergeometric function $F_2$ (Opps et al., 2005, formula (2.6)) to yields
\begin{equation}\label{sec1eq9}
H\left[\begin{array}{l}
a,b,p,\lambda    \\
\alpha,\beta,\gamma   
          \end{array}\right]=\frac{\sigma a}{4\pi}{b^{\lambda+1}\over (\lambda+1) p^\alpha} F_2(\alpha;\beta,{\lambda+1\over 2};\gamma,{\lambda+3\over 2};-{a^2\over p},-{b^2\over p})
\end{equation}
Recently, Opps et al. (2009) establish a number of new recursion formulas for the Appell hypergeometric functions $F_2$ wherein some applications to the evaluation of some generalized radiation field integrals were discussed. The purpose of the present work is to continue our investigation of finding closed form and approximation formulas for effectively computing the radiation field integrals such as equations (\ref{sec1eq1}) and (\ref{sec1eq5}). In the next section, we develop  a new approximation formula to evaluate precisely and to any desire degree of accuracy the generalized Hubbell radiation rectangular source integral (\ref{sec1eq9}). In section 3, numerical results and comparisons with previously reported values are presented. 
\section{The computation of $F_2(\alpha;\beta,{\lambda+1\over 2};\gamma,{\lambda+3\over 2};-{a^2\over p},-{b^2\over p})
$}
May be one of the most important cases regarding the computations of the radiation field integrals that capture the interest of many researchers is evaluating the integral Eq. (\ref{sec1eq1}) effectively and precisely. Some researchers were able to evaluate $h(a,b)$ using rapidly convergent series (see the original work of Hubbell et al, 1960), further Gabutti et al. (1991) investigated $h(a,b)$ in terms of its series expansions while numerical computations of this integral have been carried out by Hanak and Cechak (1978), G\"otze (1995) developed an effective method for computing the Hubbell radiation rectangular source integral, Kalla and Khajah (1997) (see also Kalla and Khajah (2000)) used Tau Method to approximate $H(a,b)$, Stalker (2001) used new convergent series for evaluating $h(a,b)$ for large $a$ and $b$ and more recently, Ezure (2005) used Haselogrove method, Guseinov et al. (2004) used binomial expansion (see also Guseinov et al. (2005)), Prabha (2006) expressed again $h(a,b)$ using some recurrence relations (see also Prabha (2007)). For a survey of various methods in computing the Hubbell rectangular source integral Eq.(\ref{sec1eq1}) and its generalization, we refer to the work of Kalla et al. 2002.
\vskip0.1true in
In this section, we given a new approximation equation that can be used to compute $H[{a,b,p,\lambda\atop \alpha,\beta,\gamma}]$ to any degree of precision and, byproduct, we can, therefore, evaluate the Hubbell radiation rectangular source integral (\ref{sec1eq1}). Our approximation expression based on the following recurrence formula for $F_2$ (see Opps et al. (2009) for detailed proof.).
\vskip0.1true in 
\noindent{\bf Theorem 1:} \emph{For $|x|+|y|<1$;~$n\geq 0;~\sigma,~\alpha_1,~\alpha_2 \in {\mathbb C};~\beta_1,~\beta_2 \in {\mathbb C} \backslash {\mathbb Z}_0^- $, the Appell hypergeometric function $F_2$ satisfies the following identity
\begin{eqnarray}\label{sec2eq1}
F_2(\sigma;\alpha_1,\alpha_2-n;\beta_1,\beta_2;x,y)&=& 
F_2(\sigma;\alpha_1,\alpha_2;\beta_1,\beta_2;x,y)\nonumber\\
&-&{\sigma~y\over \beta_2}\sum\limits_{k=1}^n F_2(\sigma+1;\alpha_1,\alpha_2-k+1;\beta_1,\beta_2+1;x,y).\qed
\end{eqnarray}
}

\noindent Writing  $F_2(\alpha;\beta,{\lambda+1\over 2};\gamma,{\lambda+3\over 2};-{a^2\over p},-{b^2\over p})$ as
 $F_2(\alpha;\beta,{\lambda+3\over 2}-1;\gamma,{\lambda+3\over 2};-{a^2\over p},-{b^2\over p})$ and apply the recurrence relation (\ref{sec2eq1}), we obtain
\begin{align}\label{sec2eq2}
F_2(\alpha;\beta,{\lambda+1\over 2};\gamma,{\lambda+3\over 2};-{a^2\over p},-{b^2\over p})&=F_2(\alpha;\beta,{\lambda+3\over 2};\gamma,{\lambda+3\over 2};-{a^2\over p},-{b^2\over p})\notag\\
&-\frac{2\alpha y}{\lambda+3}F_2(\alpha+1;\beta,{\lambda+3\over 2};\gamma,{\lambda+5\over 2};-{a^2\over p},-{b^2\over p})
\end{align} 
By means of the identity (Slater, 1960; formula 8.3.1.3)
\begin{equation}\label{sec2eq3}
F_2(\sigma;\alpha_1,\beta_2;\beta_1,\beta_2;x,y)=(1-y)^{-\sigma}{}_2F_1(\sigma,\alpha_1;\beta_1;{x\over 1-y}),
\end{equation}
we may now write Eq.(\ref{sec2eq2}) as
\begin{align}\label{sec2eq4}
F_2(\alpha;\beta,{\lambda+1\over 2};\gamma,{\lambda+3\over 2};-{a^2\over p},-{b^2\over p})&=(1+{b^2\over p})^{-\alpha}{}_2F_1(\alpha,\beta;\gamma;{-a^2\over p+b^2})\notag\\
&-\frac{2\alpha y}{\lambda+3}F_2(\alpha+1;\beta,{\lambda+3\over 2};\gamma,{\lambda+5\over 2};-{a^2\over p},-{b^2\over p}).
\end{align} 
Further, we may now regard $F_2(\alpha+1;\beta,{\lambda+3\over 2};\gamma,{\lambda+5\over 2};-{a^2\over p},-{b^2\over p})$ as $F_2(\alpha+1;\beta,{\lambda+5\over 2}-1;\gamma,{\lambda+5\over 2};-{a^2\over p},-{b^2\over p})
$ and apply the recurrence relation (\ref{sec2eq1}) again to obtain
\begin{align}\label{sec2eq5}
F_2(\alpha;\beta,{\lambda+1\over 2};\gamma,{\lambda+3\over 2};-{a^2\over p},-{b^2\over p})&=(1+{b^2\over p})^{-\alpha}{}_2F_1(\alpha,\beta;\gamma;{-a^2\over p+b^2})\notag\\
&-\frac{2\alpha y}{\lambda+3}(1+\frac{b^2}{p})^{-\alpha-1}{}_2F_1(\alpha+1,\beta;\gamma;{-a^2\over p+b^2})\notag\\
&+{2^2 \alpha(\alpha+1)y^2\over (\lambda+3)(\lambda+5)}F_2(\alpha+2;\beta,{\lambda+5\over 2};\gamma,{\lambda+7\over 2};-{a^2\over p},-{b^2\over p})
\end{align} 
After similar $n-2$ steps, we arrive at
\begin{align}\label{sec2eq6}
F_2(\alpha;\beta,{\lambda+1\over 2};\gamma,{\lambda+3\over 2};&-{a^2\over p},-{b^2\over p})=\sum_{k=0}^n (-1)^k {(\alpha)_k\over \left({3+\lambda\over 2}\right)_k}(-{b^2\over p})^k
(1+{b^2\over p})^{-\alpha-k}{}_2F_1(\alpha+k,\beta;\gamma;{-a^2\over p+b^2})\notag\\
&+{{(-1)^{n+1}(\alpha)_{n+1}y^{n+1}\over \left({3+\lambda\over 2}\right)_{n+1}}}F_2(\alpha+n+1;\beta,{\lambda+2n+3\over 2};\gamma,{\lambda+2n+5\over 2};-{a^2\over p},-{b^2\over p})
\end{align} 
where $(\alpha)_k$ denotes  the Pochhammer symbol defined, in terms of Gamma functions, by
$$(\alpha)_k= {\Gamma(\alpha+k)\over\Gamma(\alpha)}=\left\{ \begin{array}{ll}
 1 &\mbox{ if\quad $\left( k=0; \, \alpha \in {\mathbb C} \backslash \lbrace 0 \rbrace \right), $} \\
  \alpha(\alpha+1)(\alpha+2)\dots(\alpha+k-1) & \mbox{ if\quad $\left( k \in {\mathbb N}; \, \alpha \in {\mathbb C} \right)$,}
       \end{array} \right.
$$
here $\mathbb N$ being the set of {\it positive} integers. 
\vskip0.1true in
\noindent In principle, the computation of $F_2(\alpha+n+1;\beta,{\lambda+2n+3\over 2};\gamma,{\lambda+2n+5\over 2};-{a^2\over p},-{b^2\over p})$ follows the same technique and consequently, for large $n$
 \begin{align}\label{sec2eq7}
F_2(\alpha;\beta,{\lambda+1\over 2};\gamma,{\lambda+3\over 2};&-{a^2\over p},-{b^2\over p})\approx \sum_{k=0}^n {(\alpha)_k\over \left({3+\lambda\over 2}\right)_k}\left({b^2\over p}\right)^k
\left(1+{b^2\over p}\right)^{-\alpha-k}{}_2F_1(\alpha+k,\beta;\gamma;{-a^2\over p+b^2}).
\end{align} 
From which we now have for large $n$,
\begin{equation}\label{sec2eq8}
H\left[\begin{array}{l}
a,b,p,\lambda    \\
\alpha,\beta,\gamma   
          \end{array}\right]\approx\frac{\sigma a}{4\pi}{b^{\lambda+1}\over (\lambda+1)p^\alpha}\sum_{k=0}^n {(\alpha)_k\over \left({3+\lambda\over 2}\right)_k}\left({b^2\over p}\right)^k
\left(1+{b^2\over p}\right)^{-\alpha-k}{}_2F_1(\alpha+k,\beta;\gamma;{-a^2\over p+b^2})
\end{equation}
since the two-terms asymptotic expansion of the Appell hypergeometric function $F_2$ developed by L\'opez and Pagola (2008) indicate that for large $n$, the Appell hypergeometric function $$F_2(\alpha+n+1;\beta,{\lambda+2n+3\over 2};\gamma,{\lambda+2n+5\over 2};-{a^2\over p},-{b^2\over p})$$ on the right-hand side of Eq.(\ref{sec2eq6}) approach zero.
\section{Numerical results and discussion}
\noindent In order to test our approximation formula (\ref{sec2eq8}) and to show that it indeed simplify much of the numerical complexity involved in calculating the radiation field integrals such as (\ref{sec1eq1}) and (\ref{sec1eq5}), we compare, first, our approximation formula against the exact equation obtained earlier in computing 
$H\left[\begin{array}{l}
a,b,1,0    \\
\frac{1}{2},\frac{1}{2},1   
          \end{array}\right]$ 
          (Opps et al. (2009),  equation (83)),
namely, for $\sigma=1$,
\begin{align}\label{sec3eq1}
H\left[{a,b,p,1\atop \frac{1}{2},\frac{1}{2},1}\right]&={ab^{2}\over 8\pi\sqrt{p}}F_2({1\over 2};{1\over 2},1; 1,2;-{a^2\over p},-{b^2\over p})\notag\\
&={a\sqrt{p}\over 4\pi}\left[\sqrt{1+{b^2\over p}}~{}_2F_1(-{1\over 2},{1\over 2};1;-{a^2\over p+b^2})-{}_2F_1(-{1\over 2},{1\over 2};1;-{a^2\over p})\right]
\end{align}
 and
\begin{equation}\label{sec3eq2}
H\left[\begin{array}{l}
a,b,p,1    \\
{1 \over 2},{1 \over 2},1   
          \end{array}\right]=\frac{ab^2}{8\pi\sqrt{p}}\lim_{n\rightarrow\infty}\left[
\sum_{k=0}^n {({1 \over 2})_k\over \left(2\right)_k}\left({b^2\over p}\right)^k
\left(1+{b^2\over p}\right)^{-{1\over 2}-k}{}_2F_1({1\over 2}+k,{1\over 2};1;{-a^2\over p+b^2})\right].
\end{equation}
In Table 1, we reported our computation using equations (\ref{sec3eq1}) and (\ref{sec3eq2}) for different values of $a$, $b$ and $p$, the calculations were performed using MATHEMATICA software version 7. 
 It should be clear that any discrepancies may appear are due to the numerical accuracy used in computing the Gauss hypergeometric functions.
\begin{table}[h1]
\begin{center}
\caption{Comparison of the values of $H[{a,b,p,1\atop \frac{1}{2},\frac{1}{2},1}]$ for 
some values of $a$, $b$ and $p$ calculated using equations (\ref{sec3eq1}) and (\ref{sec3eq2})}
\vspace{0.1in}
\begin{tabular*}{1.0\textwidth}{@{\extracolsep{\fill}} lllll}
  \hline
\textbf{$a$} &  $b$    & $p$  & Eq. (\ref{sec3eq1}) & Eq. (\ref{sec3eq2}) \\
  \hline
0.1 & 0.2  & 0.5  & 0.000~219~698~305~361~162~27&0.000~219~698~305~361~161~92 \\
0.1 & 0.5  & 0.5  & 0.001~259~518~891~975~573~3& 0.001~259~518~891~975~572~7 \\
0.2 & 0.2  & 2.0  &0.000~222~868~191~957~851~78 &0.000~222~868~191~957~853~9 \\
0.2 & 1.0  & 2.0  &0.005~038~075~567~902~293 &0.005~038~075~567~902~291  \\
0.5 & 0.5  & 0.5  &0.005~794~884~270~704~959~5 &0.005~794~884~270~704~952~5 \\
0.5 & 1.0  & 2.5  &0.011~293~885~774~813~334~5 &0.011~293~885~774~813~332 \\
\hline
\end{tabular*}
\end{center}
\end{table}
In Table 2, we compared our calculated values with those obtained by Guseinov and Memedov (2005) and with those obtained by Galu\'e et al. (1994).  It should be clear that equation (\ref{sec2eq8}) can be used for arbitrary values of the parameters 
$a$, $b$, $p$ and it is not restricted to any particular range of parameter values.

\begin{table}[h1]
\begin{center}
\caption{The values of $H[{a,b,p,1\atop \frac{1}{2},\frac{1}{2},1}]$ integrals for $\sigma=1$ 
and some values of $a$, $b$ and $p$ obtained from Eq. (\ref{sec2eq8}), Guesinov and Mamedov (2005)
 and Galu\'e et al. (1994).}
\begin{tabular*}{1.1\textwidth}{@{\extracolsep{\fill}} llllll}
  \hline
\textbf{$a$} &  $b$    & $p$  & Eq. (\ref{sec2eq8}) & Guseinov \& Mamedov &
Galu\'e et al. \\
~ &  ~    & ~  & ~ & ~(2005)&
(1994)\\
  \hline
0.1 & 0.2  & 0.5  & 0.000~219~698~305~361~161~92& 0.000~219~698~305~352~979&
0.000~219~698~31\\
0.1 & 0.5  & 0.5  & 0.001~259~518~891~975~572~7&  0.001~259~518~892~096~95&
0.001~259~518~9\\
0.2 & 0.2  & 2.0  & 0.000~222~868~191~957~853~9& 0.000~222~868~191~569~00&
0.000~222~868~19\\
0.2 & 1.0  & 2.0  & 0.005~038~075~567~902~291& 0.005~038~075~568~393~22&
0.005~038~075~6\\
0.5 & 0.5  & 0.5  & 0.005~794~884~270~704~952~5& 0.005~794~884~271~955~31&
0.005~794~884~3\\
0.5 & 1.0  & 2.5  & 0.011~293~885~774~813~332& 0.011~293~885~774~293~0&
0.011~293~886\\
\hline
\end{tabular*}
\end{center}
\end{table}
\vskip0.1true in
In Table 3, we report our numerical computation of the Hubbell  rectangular source integral (\ref{sec1eq1}) for $\sigma =p=1$ as well for some other values of $p$ using equation (\ref{sec2eq8}), or simply
 \begin{equation}\label{sec3eq3}
H\left[\begin{array}{l}
a,b,p,0    \\
1,\frac{1}{2},\frac{3}{2}   
          \end{array}\right]\approx\frac{\sigma ab}{4\pi~{p}}\sum_{k=0}^n {k!\over \left({3\over 2}\right)_k}\left({b^2\over p}\right)^k
\left(1+{b^2\over p}\right)^{-1-k}{}_2F_1(k+1,\frac{1}{2};\frac{3}{2};{-a^2\over p+b^2})
\end{equation}
In the same table, we also compared our results with the earlier numerical values obtained of Guseinov and Mademov (2005) and  
Galu\'e et al (1994). 
\begin{table}[h]
\begin{center}
\caption{The comparative values of $H[{a,b,p,0\atop 1,0.5,1.5}]$ integral from Eq.(\ref{sec2eq8}), 
Guseinov and Mamedov (2005) and Galu\'e et al (1994)}
\vspace{0.1in}
\begin{tabular*}{1.0\textwidth}{@{\extracolsep{\fill}} llllll}
  \hline
\textbf{$a$} &  $b$    & $p$  & Eq. (\ref{sec2eq8}) & Guseinov \& Mamedov &
Galu\'e et al. \\
~ &  ~    & ~  & ~ & ~(2005)&
(1994)\\
  \hline
0.1 & 0.1  & 0.5  & 0.001~570~716~369~171~686& 0.001~570~716~369~157~32&
0.001~570~716~4\\
0.1 & 0.5  & 1.0  & 0.003~678~199~808~681~331& 0.003~678~199~808~778~47&
0.003~678~199~8\\
0.2 & 0.5  & 2.5  & 0.003~067~148~756~523~266& 0.003~067~148~756~507~53&
0.003~067~148~8\\
0.2 & 0.8  & 2.0  & 0.005~758~600~701~534~181& 0.005~758~600~701~357~22&
0.005~758~600~7\\
0.5 & 0.5  & 1.0  & 0.017~188~506~077~049~23& 0.017~188~506~077~717~6&
0.017~188~506\\
0.6 & 0.5  & 1.0  & 0.020~067~469~440~496~68& 0.020~067~469~441~718~5&
0.020~067~469\\
0.8 & 0.6  & 2.8  & 0.012~248~693~964~171~94& 0.012~248~693~963~979~3&~\\
1.0 & 0.8  & 4.2  & 0.013~484~796~561~457~52&0.013~484~796~561~864~6&~\\
0.5 & 2.0  & 5.4  & 0.012~012~547~384~013~146&0.012~012~547~385~362~9&~\\
0.8 & 2.6  & 7.5  & 0.017~255~112~588~899~273&0.017~255~128~853~560~6&~\\
\hline
\end{tabular*}
\end{center}
\end{table}
\noindent It is worth noting an important feature of the approximation formula (\ref{sec2eq8}) is that it is \emph{self-adjusting}; that is, if
\begin{equation}\label{sec3eq3}
|H_{n+1}-H_n|\not<\epsilon
\end{equation}
where $\epsilon$ is the desire accuracy then $n$ should be increased to reach the required accuracy. Here, $H_n$ refer to the right-hand side of equation (\ref{sec2eq8}).  
\vskip0.2 true in

\noindent{\bf Acknowledgment.} The present investigation was supported, in part, by the Natural Sciences and Engineering Research
Council of Canada under Grant  GP249507.

\end{document}